\documentclass[12pt, letterpaper]{article}

\usepackage{amsmath,amssymb}
\usepackage{cite}
\usepackage{fancyhdr}
\usepackage[top=1in, bottom=1in, left=1in, right=1in]{geometry}
\usepackage[dvipdfm, dvips]{graphicx}
\usepackage{hyperref}

\numberwithin{equation}{section}

\begin{document}


\setcounter{page}{0}
\date{}

\lhead{}\chead{}\rhead{\footnotesize{RUNHETC-2012-07\\SCIPP-12/06}}\lfoot{}\cfoot{}\rfoot{}

\title{\textbf{The Top $10^{500}$ Reasons Not To Believe in the String Landscape\vspace{0.4cm}}}

\author{Tom Banks$^{1,2}$ \vspace{0.7cm}\\
{\normalsize{$^1$NHETC and Department of Physics and Astronomy, Rutgers University,}}\\
{\normalsize{Piscataway, NJ 08854-8019, USA}}\vspace{0.2cm}\\
{\normalsize{$^2$SCIPP and Department of Physics, University of California,}}\\
{\normalsize{Santa Cruz, CA 95064-1077, USA}}}

\maketitle
\thispagestyle{fancy}

\begin{abstract}
\normalsize \noindent
The String Landscape is a fantasy.  We actually have a plausible landscape of minimally supersymmetric $AdS_4$ solutions of supergravity modified by an exponential superpotential.  None of these solutions is accessible to world sheet perturbation theory.  If they exist as models of quantum gravity, they are defined by conformal field theories, and each is an independent quantum system, which makes no transitions to any of the others.  This landscape has nothing to do with CDL tunneling or eternal inflation.

A proper understanding of CDL transitions in QFT on a fixed background dS space, shows that the EI picture of this system is not justified within the approximation of low energy effective field theory.  The cutoff independent physics, defined by the Euclidean functional integral over the 4-sphere admits only a finite number of instantons.  Plausible extensions of these ideas to a quantum theory of gravity obeying the holographic principle explain all of the actual facts about CDL transitions in dS space, and lead to a picture radically different from eternal inflation.

Theories of Eternal Inflation (EI) have to rely too heavily on the anthropic principle to be consistent with experiment. Given the vast array of effective low energy field theories that could be produced by the conventional picture of the string landscape one is forced to conclude that the most numerous anthropically allowed theories will disagree with experiment violently. Ideas about terminating EI by a Census Taker's Hat, can resolve some of these issues because they claim that the late time dynamics of EI is determined by a single eigenvalue of a CDL generated Markov process.  That eigenvalue in turn is dominated by a single, highly supersymmetric vacuum state, which is averred to be hostile to life of our type.  Other components of the dominant eigenvector arise in low orders of perturbation theory in transitions out of this state, and one asserts that the lowest order contribution supporting life of our type is unique and will predict the physics we observe.

This paper is based on lectures given at UCSC in the spring of 2012, and at the Cargese Summer Institute in June 2012.
\end{abstract}


\newpage
\tableofcontents
\vspace{1cm}

\section{Introduction}

Virtually all of the work, which attempts to make a connection between string theory and the real, non-supersymmetric, world, centers on the concept of the string landscape.  Although I worked extensively on this subject in the mid 1990s\cite{tblandscape}, and was one of those responsible for the anthropic argument that lends the only bit of phenomenological credence to these ideas\cite{tblindecc}, I have, since about 1999, argued that there is no actual scientific evidence for the landscape, that the whole concept is based on faulty ideas about quantum gravity, and that the naive phenomenological success of the landscape in fact masks a much larger phenomenological failure, which is usually ignored by the partisans of this picture.  I spent some time trying to refute these ideas in the early years of the 21st century.  This paper is an update on that refutation.  It contains the following key ideas

\begin{itemize} 

\item The landscape idea is based on the notion of a quantum effective potential, for whose existence there is in fact no scientific argument in the formalism of string theory.  Indeed, {\it all} evidence, from both perturbative and non-perturbative string theory, points to a quite different notion, namely that {\it globally different space-times correspond to different quantum Hamiltonians, rather that different states of a single quantum theory. } The extant evidence also suggests that asymptotically flat space, and asymptotically anti-de Sitter (AdS) space  of curvature much smaller than the string scale, only exist in conjunction with exact supersymmetry\footnote{Supersymmetric conformal field theories with large radius AdS duals, perturbed by SUSY violating {\bf relevant} operators, can be viewed as non-supersymmetric objects embedded in a space-time, which is supersymmetric in the bulk.}.  I'll argue in particular that the Fischler-Susskind mechanism points to singular time dependent solutions of gravity, rather than controllable perturbation expansions.  Indeed, even in a class of examples where effective field theory reasoning leads to a scenario with a stable super Poincare invariant model, modified slightly from a world sheet model by  perturbative corrections, we have no method for constructing a systematic perturbation expansion using world sheet techniques. The String Landscape should really be called the Supergravity Landscape.  It's main achievement is the establishment, with a high degree of plausibility, of a large class of large radius supersymmetric AdS models with cosmological constant that may be tuned very small in string units by varying fluxes.  

\item The idea of {\it flux compactification}\cite{fluxcomp} on which much of the landscape theory rests, is haunted by a classic no-go theorem\cite{tenintofour}. The attempt to get around this is based on the idea of {\it orientifold planes}, and in many cases one can argue that these do not solve the problem.  There is, in my opinion, only one well motivated set of flux compactifications, corresponding to $AdS_4$ space-times, with AdS radius parametrically larger than that of the compact dimensions.  However, I do not want to claim that these are the only flux compactifications, but rather that any other collection of compactifications will have qualitatively similar properties. The favored compactifications arise from a small modification of classical solutions of Type IIB Supergravity (SUGRA), where the Type IIB coupling $\tau$ parametrizes the elliptic fiber of a Calabi-Yau fourfold, solutions commonly known as F-theory.  The classical solutions violate both R symmetry and supersymmetry, but have exactly vanishing cosmological constant.  In certain limiting cases, they admit a tree level world sheet theory, but the perturbation expansion around those cases has Fischler-Susskind divergences and leads to a time dependent and asymptotically singular geometry, for which the perturbation expansion breaks down.  However, when one uses the freedom of choosing fluxes, first indicated by Bousso and Polchinski\cite{bp}, one can tune the R-breaking to be very small.  In these cases, a plausible modification of the superpotential can lead to a stable, supersymmetric AdS solution.   We emphasize that the procedure of ``uplifting" this solution to a meta-stable de Sitter space is completely unjustified.  Apart from technical problems about the meaning of uplifting in effective field theory, one has to admit that the ``theory" of meta-stable dS spaces and eternal inflation, is wildly different from the CFT which might serve as the quantum definition of the F-theory AdS space.  Thus, to claim that one can derive one as a small perturbation of the other is, on the face of it, absurd.
The F-theory flux compactification landscape (and other constructions, to the extent that they are plausible) is really a collection of superconformally invariant $2+1$ dimensional field theories, between which there are no dynamical transitions.  Each is a unitary theory of quantum gravity.
They thus stand as 
$\sim 10^{500}$ counterexamples to the idea of a connected landscape of meta-stable states.

\item In quantum field theory, one can verify the validity of the claim that one meta-stable minimum of an effective potential is a state in a model defined by the stable minimum, by constructing scattering experiments, which enable one to create that state over an arbitrarily large region of space.  In models of gravity this {\it always} fails, and any attempt to do this leads to the creation of black holes instead.  In particular, if the meta-stable state was a dS space, one can show that black hole production always sets in before the external observer can induce a local region of space to undergo exponential expansion. 

\item Coleman De Luccia (CDL) tunneling is the classic mechanism for exploring multiple minima of the potential.  Without gravity, bubble nucleation provides a way for a meta-stable minimum of the potential to decay into the stable state.   Once gravity is taken into account, this occurs only if the "true" minimum has non-negative c.c. .  In these cases, the CDL process represents a decay. By contrast, if  the true c.c. is negative, the final state of tunneling is a singular space-time in which the fields do not stay in the basin of attraction of the minimum. When the false minimum has vanishing or small c.c., of either sign, the space of potentials breaks up into two classes\cite{abj}.  In the first class, no tunneling occurs if the c.c. is non-positive, and for positive c.c. the CDL process is not a decay, but a Poincare recurrence in which a finite entropy system violates the second law and makes a brief sojourn into a very low entropy state.  For potentials of the second class, which we say are {\it below the Great Divide}, the decay of the vacuum does not imply decay of excitations of the vacuum.  The collision of a bubble and a black hole leaves the interior of the black hole unscathed unless the bubble is swallowed by the black hole.  In the latter case, the world external to the black hole sees no decay, and much of the interior is causally disconnected from the CDL singularity.  It is unclear whether potentials below the great divide occur in real theories of quantum gravity.   They play a crucial role in the FRW/CFT\cite{frwcft} theory of eternal inflation.  That theory is based on CDL bubbles whose true minimum has zero c.c. , and we will make some brief comments on it below.

\item Matrix Theory\cite{bfss} and AdS/CFT\cite{aharony} both provide enormous amounts of evidence that different space-time backgrounds correspond to different models of quantum gravity, each with its own Hamiltonian.  It is similar to the plethora of models of quantum field theory, with the following difference.   All well established models with AdS radius of curvature much larger than the string scale, so that effective field theory methods are valid for at least some observables, are exactly supersymmetric in the bulk of space-time.  I'll briefly review the salient points below. The relation between the asymptotics of space-time and the Hamiltonian of quantum gravity follows from very general principles:  Energy is always defined at infinity in general relativity.  The deep UV, which is the defining regime for every known quantum theory with an infinite dimensional Hilbert space, is plausibly dominated by black hole physics.   The high energy spectrum of black holes depends crucially on the long wavelength properties of space-time, and in particular, on the value of the c.c. .

\item Weinberg's calculation\cite{weinberg}, following the suggestions of the author and Linde\cite{tblindecc}, of the anthropic bound on the c.c. , is the phenomenological success, which drives the arguments for the landscape/eternal inflation picture.  However, if too many parameters of low energy physics fluctuate in the landscape, as seems inevitable if the string landscape exists, then there are more phenomenological failures than successes.  

\end{itemize}

\section{Barriers to Potentials}

In tree level string theory, space-time has an asymptotically flat or AdS factor, and correlation functions on the conformal boundary of space-time are computed as expectation values of products of vertex operators in a world sheet conformal field theory.  The vertex operators, which have fixed Mellin weights in AdS space\cite{Mellin} or fixed momentum in Minkowski space, have world sheet dimension two and their one point functions on spherical world sheets all vanish.  In certain kinematic circumstances (large AdS radius, in string units, small values of the Mellin invariants/momenta) the same scattering amplitudes can be computed approximately in terms of a bulk effective Lagrangian for space-time fields.  The vanishing one point functions imply that the background field configuration is at a stationary point of the scalar field potential $V(\phi )$ in the effective field theory.  

If the tree level theory is tachyon free, then potential divergences in the string loop expansion, correspond to the possibility that one point functions for some of the massless scalars exist on higher genus Riemann surfaces.   The integral over moduli of those surfaces has a divergence that corresponds precisely with the exchange of a zero momentum massless propagator between a tadpole and a lower genus scattering amplitude\cite{fischlersusskind}.  Fischler and Susskind\cite{fischlersusskind} argue that this divergence can be canceled exactly, by allowing the background to be time dependent.   In cases where the low energy physics is described by a space-time effective Lagrangian, the same time dependent background is a solution of a modified set of Lagrange equations, for a Lagrangian with a potential.  

The actual time dependent solution is linear in time in the perturbative approximation.  We can try to extrapolate to larger times by looking at non-linear solutions of the space-time effective action.  However\cite{bd} these are always singular in either the past or the future, as a consequence of the Hawking Penrose theorems.   Scattering amplitudes no longer make sense and neither the low energy effective field theory, nor the string loop expansion are under control. Thus, there is in fact no way to turn the Fischler-Susskind mechanism into a systematic perturbation expansion.

It's quite remarkable that this is even true of a case where low energy effective field theory indicates that the resolution of the divergence problem is simple and stable.  In some supersymmetric compactifications of the heterotic string, there are $U(1)$ gauge bosons that appear to have a gauge-gravitational anomaly.  This leads to an apparent one loop tadpole for the D term of the vector multiplet, which would break SUSY. The authors of \cite{dsww} have shown that the dilaton superfield can always be given a $U(1)$ transformation law to cancel the anomaly.  In low energy effective field theory this implies that there is a Higgs mechanism in which a combination of the dilaton and another charged field get a VEV, leaving over a massless chiral superfield whose expectation value can serve as a tunable coupling parameter.   This is an example where the Fischler-Susskind analysis seems to indicate the existence of a weakly coupled supersymmetric string model in asymptotically flat space, with a weakly Higgsed $U(1)$ gauge field.  Nonetheless, the past $25$ years have not led to a systematic world sheet perturbation expansion for this system.  We are left only with a plausible, non-renormalizable low energy field theory, whose perturbation expansion contains all the usual ambiguities.

Often, in discussions of the string landscape, some effort is made to ensure that the moduli are stabilized in a regime where the string coupling is weak.  This is supposed to give the argument more weight, since without a worldsheet expansion we are simply working with a non-renormalizable effective field theory.  Our discussion of the Fischler-Susskind method shows that, at least at our current level of understanding, world sheet methods cannot find new minima that one would like to associate with quantum corrections to the space-time effective potential.  They give no prescription for defining such a potential.  At most they give us evidence for singular time dependent solutions in which one of the tree level moduli moves down the slope of the effective potential.

I want to emphasize that the arguments above do not in any way affect the use of space-time effective actions to prove the exact or perturbative stability of certain string models.  The typical form of such an argument is:  If this model is unstable, it is because of a tadpole for some massless scalar field.
Supersymmetry, sometimes combined with other symmetries, forbids any term in an effective action which could generate such a tadpole, or shows that any such term is exponentially small when the string coupling is small.
Thus, effective actions may be used to describe low energy scattering in a given string model, or to prove the stability of such a model (at least to all orders in perturbation theory).  They cannot be used to give a systematic perturbative correction to the asymptotic space-time of the model itself.

\section{The Landscape of F-theory}

Perturbative  string theory provides us with a large number of space-times that have ${\cal N} = 1$ Super-Poincare invariance in four Minkowski dimensions. All known examples have continuous moduli, which appear as chiral superfields in the low energy effective action.  There are similar examples in 11D SUGRA compactified on manifolds of $G2$ holonomy, and in the limiting case of three dimensional compactifications of 11D SUGRA,  which is called F-theory on a Calabi-Yau fourfold.   F-theory models are based on solutions of Type IIB SUGRA in which the axion-dilaton field $\tau (y)$ varies over a three dimensional complex base space with coordinates $y$, to make an elliptically fibered $CY_4$ manifold.   

The space of moduli of all of these constructions is non-compact, and recent results\cite{komseib} show that compact moduli spaces can exist only at the Planck scale, if at all. In particular, all constructions are perturbation theories based on the assumption that at least one of the moduli is near infinity in the
moduli space.  For perturbative string theory, this is the dilaton or string coupling modulus, while for M and F-theory compactifications it is the volume modulus of the $G2$ manifold or the base of the elliptically fibered $CY_4$\footnote{We focus attention on F-theory compactifications only because this case has a relatively straightforward description in terms of generically smooth classical solutions.  There are many other classes of solutions, which lead to minimally supersymmetric $AdS_4$ compactifications, and I'm sure that in at least some of them, one can argue that the AdS radius can be much larger than the internal radius, and both larger than the Planck scale.  I have not been able to do a general survey of the vast literature on this subject, and in the end it would do no good.
All such models will suffer from similar criticisms.  The only way to find a perturbative world sheet theory of $AdS_4$ with non-perturbatively stabilized moduli, would be the analog of what happens in $D1-D5$ realizations of $AdS_3 /CFT$, where the string coupling is really a discrete variable.
Chern-Simons models coupled to free conformal matter would seem like a good starting point for such a search.}

Discrete shift symmetries of the real part of the large modulus guarantee that any superpotential for the moduli is exponentially small when the modulus is in its asymptotic region.   Thus, one expects a stable perturbation series in powers of the large modulus.  In weakly coupled string theory we have unambiguous rules for constructing this perturbation expansion, while in the M and F theory cases we are doing a low energy effective field theory expansion, and each order introduces more unknown coefficients.

Flux compactifications\cite{flux}\cite{fluxcomp} were introduced for two purposes.  The first is that, in the presence of fluxes, most of the moduli can be fixed at leading order in the expansion.  The second is that, on a topologically complicated manifold, with many relatively large fluxes, one has a source of large numbers, which might stabilize the dilaton/volume moduli at large values where the initial approximation is valid.  Finally, Bousso and Polchinski\cite{bp} have argued that the presence of many fluxes can provide a mechanism for tuning the cosmological constant to zero.   In practice, this is achieved by finding supersymmetric solutions with small negative c.c., and then ``uplifting them" to find dS solutions.   In my opinion, the uplifting procedure is invalid, so I will begin by describing the search for supersymmetric vacua with small negative c.c. .   I will restrict attention to the F-theory examples, which appear to be the most robust and plausible.

The F-theory flux compactifications are well described in Chapter 10 of the textbook by Becker, Becker and Schwarz\cite{bbs}.  One begins with a complex 3-fold which is conformal to a $CY_3$.   The space-time metric
is 
$$ ds^2 = \Delta (y)^{-1} (dx^{\mu})^2 + \Delta (y)^{1/2} g_{ij} dy^i dy^j ,$$ and the axion-dilaton is $\tau (y) $.  

Flux compactifications lead to a four dimensional effective field theory in which the Kahler moduli of the $CY_3$ are chiral superfields.  We will restrict attention to models with a single Kahler modulus, $\rho$, but the general pattern is similar.  Realistic models will also contain a four dimensional supersymmetric gauge theory, in which the standard model is embedded.
The superpotential is $$ W = W_0 + \sum_{n > 0} a_n e^{i n a \rho} .$$ Im $\rho$ is positive and there is a discrete shift symmetry $\rho \rightarrow \rho + \frac{2\pi}{a} $.  The constant $W_0$ is generically non-zero, and vanishes only if the model preserves a discrete R-symmetry.   Such symmetries are rare\cite{dinesun} but do exist for some flux choices.   This symmetry might act on $\rho$ by a discrete shift, in which case some of the exponential terms in the superpotential are allowed.

In the R symmetric case with no allowed exponentials, we have an effective field theory argument for a stable, exactly supersymmetric model of four dimensional quantum gravity\cite{bd3}.  It will have a super-Poincare invariant S matrix, and an exact modulus, $\rho$.   In certain orientifold limits, it will have a worldsheet perturbation expansion, as well as an effective field theory expansion, if the fluxes can be tuned so that the constant value of the string coupling is small.  If $\rho$ transforms under the R-symmetry, so that exponential terms in the superpotential are allowed, we expect an instability that can be seen neither in the large Im $\rho$ expansion, nor in world sheet perturbation theory, in cases where the latter is applicable.

If there is no R symmetry, so that $W_0$ is non-vanishing, the situation is quite different.  It turns out that one can only do a plausible analysis when $W_0$ is much smaller than $m_P^3$ , where $m_P$ is the four dimensional Planck mass.  Here is where the Bousso-Polchinski counting arguments come in.  One argues that $W_0$ is a sum of many complex terms, if the topological complexity of the manifold, (the third Betti number $B_3$ ), is large.  Each term can take on a few discrete values, by changing the flux on a particular three cycle.  By tuning the fluxes one argues one can find a small number of solutions with c.c. as small as $10^{ - B_3} m_P^3 $.
Estimates of the maximal $B_3$ for $CY_3$ manifolds are of the order $500$.  Kachru et. al. \cite{kklt} have found explicit examples where the cancelation is good to one part in $10^{-4} $ or $10^{-5}$.

If we neglect the terms in the superpotential exponential in $\rho$ we find that supersymmetry is broken unless Im $\rho = \infty$, but the Kahler potential in the SUGRA Lagrangian takes on the {\it no scale} form
$$K = 3\  {\rm ln\ Im}\ \rho ,$$ which gives a vanishing c.c. for all values of $\rho$.  There is no potential, despite large breaking of SUSY.  Note however that {\it any} correction to $K$, subleading for large Im $\rho$, will ruin this cancelation and give rise to a potential which is positive and tends to zero for $\rho \rightarrow \infty$.  The corresponding classical solution is a $4D$ FRW cosmology, which generically has a Big Bang or Big Crunch.  To be more precise, the Fischler-Susskind calculation will give a linear time dependence to moduli, and one can try to re-sum that secular breakdown of perturbation theory by solving the non-linear classical equations, with the corrected Kahler potential , which provides a potential that fits the linearized F-S equations.  It is these classical equations which have Big Bang or Crunch solutions.  Note that there is no way in which the Fischler-Susskind algorithm could be sensitive to the exponential correction to the super-potential which, for tuned values of $W_0$, can give rise to supersymmetric AdS solutions with large radius of curvature.

The ``String Landscape" is thus a landscape of solutions of a slightly modified Type IIB supergravity Lagrangian, which has no real connection to perturbative string theory.  It is based on a self consistent approximation in which the volume of compact dimensions is large in Planck units, and one has enough freedom to tune the radius of the AdS space to be much larger than that of the internal manifold.
As an effective field theory, Type IIB SUGRA is non-renormalizable, so higher order corrections will introduce ambiguities in the expansion in large internal radius.

It would seem that the only way to go beyond the EFT treatment would be to find CFT duals for these solutions. 
Somewhat surprisingly, very little work\cite{eva} has been done on the search for such duals.

In a typical discussion of the landscape, the next step would be to discuss the process of ``uplifting" an AdS vacuum with small negative c.c. to get a meta-stable dS space.  There are many technical problems associated with the uplifting process, but I think the overwhelmingly important fact is that all proposals for a {\it theory} of a meta-stable dS landscape are infinitely far away from the boundary CFT which gives the proper definition of quantum gravity in AdS space.   It seems wrong on the face of it to claim that the dS landscape is a small perturbation of the theory of an AdS space based on a nearby stationary point of the potential.  The theory of meta-stable states and tunneling in the presence of gravity was initiated by Coleman and De Luccia.  It is extremely subtle, and there are many sub-cases.  We will discuss it in the next section.  One of its many lessons is that AdS space is {\it never} the result of tunneling from some meta-stable point on the effective potential.

The summary of the message of this and the previous section is that, given standard estimates of the degree of topological complexity to be expected from Calabi-Yau 3 and 4 folds, there is a plausible argument that there are $10^{500}$ models of quantum gravity in $AdS_4$ spaces with minimal SUSY.  Using a version of the arguments of Bousso and Polchinski, one can argue that some of these models will have AdS radius much larger than the radius of compact dimensions.   These models are solutions of Type IIB SUGRA modified by a superpotential exponential in the Kahler moduli of the internal manifold, and those moduli are stabilized at a radius much larger than Planck scale, so that the SUGRA description is justified.

All these models are low energy effective field theories, and there is no systematic way to compute unambiguous higher order corrections to the long wavelength approximation. The generic model in this class is an F-theory model, and has no world sheet interpretation.  In certain limits the models become orbifolds, where the effective field representing the string coupling $\frac{1}{{\rm Im}\  \tau}$ is small almost everywhere.  However, we've argued that in those limits a world sheet expansion would lead to an uncontrolled Fischler-Susskind cosmology, rather than to the stable solution of the effective field theory.  Thus, the only way to connect these solutions to ``string theory" would be to find the CFTs to which they should be dual.   This has not been done.  It is a worthwhile project and it would show that each of these different solutions represents a completely different quantum system, with different high energy density of states, and an inequivalent operator product algebra.  

Thus, these AdS models are, if confirmed, $10^{500}$ or more reasons {\it not} to believe in a connected landscape of meta-stable vacuum states, which are all part of the same theory.   In the next section we will study the only theoretical tool we have for exploring the validity of the landscape idea.  The results are not without ambiguity, but one thing is clear from them.   In {\it no} case is an AdS state connected by tunneling to a meta-stable or stable dS space, or part of a possible model of Eternal Inflation.  The idea that EI is just a ``small uplifting" of a stable AdS background is simply wrong. Instead, the $10^{500}$ AdS models, if they exist are independent unitary conformal field theories, each of which defines a different model for quantum gravity in a space-time of relatively low curvature.

\section{Coleman De Luccia (CDL) Tunneling}

Much of the theory of the landscape and eternal inflation is based on the Coleman De Luccia analysis\cite{cdl} of tunneling in the presence of gravity.  Much of the actual content of that analysis is ignored in much of the literature on eternal inflation, so it's worth summarizing the CDL results, along with a cogent analysis of their meaning in the case of tunneling from de Sitter space, due to Brown and Weinberg\cite{BW}.

We begin by analyzing the case of tunneling from a 4 dimensional space-time with zero cosmological constant.  We choose a model with a single scalar field, but the results are more general.  The potential has the form $V = -\mu^4 u(\phi / M) $, and we introduce dimensionless field and coordinate variables, expressing all dimensions in terms of $\frac{\mu^2}{M}$.  We will call the dimensionless field variable $f$, and introduce $\epsilon^2 \equiv \frac{M^2}{3 m_P^2} $.  The action is then of order $\frac{M^4}{\mu^4}$ and
 the semi-classical approximation is (nominally) valid when this ratio is large. We introduce the metric ansatz

$$ds^2 = - dz^2 + a^2 (z) d\Omega_3^2 ,$$ and require $f$ to be a function of $z$ .  At infinity, we must have $a(z) \rightarrow z$.   At some point, which we conventionally call $z=0$, $a(z)$ must vanish linearly $ a \rightarrow z $ in order to avoid a conical singularity.  This is the center of the bubble.  $f$ must approach the false Minkowski vacuum value where $u(0) = 0$, and we have conventionally chosen this to be at $f = 0$.  The Euclidean equations of motion are

$$f_{zz} + 3\frac{a_z}{a} f_z + u^{\prime} (f) = 0 .$$ $$ a_z^2 = 1 + \epsilon^2  a^2 (\frac{f_z^2}{2} + u(f) ) .$$  These resemble the equations for an FRW cosmology with negative spatial curvature, with a scalar field propagating in the upside down potential $u$.  Note that the Euclidean manifold has sections of positive curvature, but that $z$ is a positive signature variable.

The analysis of the instanton in this analog cosmology is useful both because it makes the results intuitive, and because the same equations arise in the analysis of the classical motion after tunneling.  To avoid a singularity at $z=0$, we must have $f_z (0) = 0$, so our only free boundary condition is $f(0)$.   By convention, we put the true minimum at $f_T > 0$, so in order to qualify as in instanton, we must have $f(0) > 0$ , and $f(z)$ must asymptote to $0$ as $z \rightarrow \infty$.

When $\epsilon = 0$, it's clear that a solution always exists.  If $f(0) $ is near the minimum of $u$, then, since the ``energy" $\frac{f_z^2}{2} + u$, is always decreasing under the influence of friction, the solution undershoots.  On the other hand, if we take $f(0)$ very near the maximum $f_T$ of $u$, then it hovers there for a long time, while the friction term goes to zero like $1/z$.  The solution clearly overshoots the maximum at $f = 0$.  By continuity there is a solution which asymptotes to $f = 0$.  As we vary the potential, this solution exists as long as $u(f_T) > 0$.  In the limit of degenerate minima the instanton approaches a static domain wall between the two minima.

When $\epsilon > 0$ a new situation arises.  For large $a$ the gravitational correction is always important, and if $f$ has not yet crossed into the regime
where $ u > 0$, on the side of the minimum close to $f = 0$, then $a_z$ vanishes and then changes sign.   $a$ returns to zero, but this time $f_z$ doesn't vanish simultaneously. The solution is singular and never asymptotes to flat space.  The strategy of waiting on the true vacuum side of the minimum until the friction goes away, does not work.   It's easy to see that when $\epsilon << 1$, this singularity does not occur.  The transit time is $o(1)$ but the time it takes for $\epsilon a $ to be $\sim 1$ is $o(1/\epsilon) $.   The transition occurs for some $\epsilon$ of order one, unless $u(0) \ll 1$, in which case the singularity can occur at small $\epsilon$.  It's possible to show that, in the space of all potentials with a Minkowski minimum, the boundary surface between potentials with and without instantons has co-dimension one\cite{abj}.  This surface is called The Great Divide.

Potentials above The Great Divide have absolutely stable Minkowski minima, with a Positive Energy Theorem\cite{Schoenyau}.  Right on the great divide we have domain walls.  Domain walls have a gravitational field, which leads to a jump in the c.c. as they are crossed, so when $\epsilon$ is non-zero, the transition occurs when the true minimum has negative c.c. .

The Lorentzian evolution of the bubble, after nucleation, has the same characteristics, inside the bubble as the Euclidean analog cosmology in the case where no instanton exists.  The crucial difference is that we no longer have the freedom to choose the initial condition for $f$ at the point where $a = 0$.  {\it Thus, for generic potentials, whenever a CDL decay to a region of negative potential $V$ exists, it leads to a singular cosmology.  No FRW model can settle down to an AdS minimum.  Indeed, it is misleading to say that the instanton decays to the basin of attraction of the negative minimum.  In the singular cosmology, $\dot{a}/a$ is negative, the field feels ``anti-friction" and its energy goes to infinity.  It does not stay localized on any portion of the potential, but explores more and more of the phase space of the scalar.}. 
AdS minima are {\it never} final states of CDL tunneling processes, and are only initial CDL states below the great divide.

Potentials below the Great Divide have problematic aspects, when we consider the quintessentially gravitational process of collision between an expanding bubble of true minimum and a black hole.  Clearly, if the bubble is smaller than the horizon size when the collision occurs, then the black hole swallows the bubble and the universe is saved from decay!  This is in marked contrast to what happens in quantum field theory, where one {\it catalyzes} vacuum decay by increasing the energy density in a local region.
Furthermore, even if the bubble is much larger than the black hole at the time of collision, the expanding bubble does not penetrate the interior of the black hole, because it expands only at the speed of light.  Note that both the interior of the bubble, and that of the black hole become singular in finite proper time.  For the bubble, that proper time is a fixed microscopic scale, independent of the size of the bubble as viewed from the exterior.  For the black hole the time to the singularity scales with the horizon radius.  So we can have a situation in which an observer faced with the prospect of being engulfed by a collapsing bubble, can live longer\footnote{and, according to the Covariant Entropy Bound, do many more experiments,} by jumping into the black hole.  

In quantum field theory, the final state of a decay by bubble nucleation is a quiescent true vacuum. This is independent of the state of excitation of the false vacuum.  In the presence of gravity, if the false vacuum c.c. vanishes and the true minimum has negative c.c., the final state of a CDL decay depends crucially on the initial state of excitation of the false vacuum.  Black holes can swallow bubbles, and the generic time-like infinity consists of a collection of causally disconnected singular regions.  It's clear that low energy effective field theory breaks down completely for these systems, and one may legitimately wonder whether they correspond to {\it any} well defined theory of quantum gravity.

This picture is substantially unchanged if we move the false minimum to slightly negative c.c. .  As has become familiar from the AdS/CFT correspondence, a positive curvature minimum in AdS space leads to two power law behaviors at infinity, one of which grows, and the other of which is a normalizable fluctuation corresponding to an excited state of the AdS vacuum.  The tuning of $f(0)$, discussed above for flat space, corresponds to tuning the coefficient of the non-normalizable fluctuation to zero.  Thus, the instanton represents the creation of a normalizable state in the bulk AdS field theory.  The Great Divide still exists, and is related to positive energy theorems in AdS space\cite{horowitz}.  Note however that if the false minimum c.c. is sufficiently large and negative, {\it all} potentials are below the great divide.

There is a difference between the $\Lambda = 0$ and $\Lambda < 0$ cases when it comes to collective coordinates.  Both space-times have non-compact asymptotic symmetry groups and the instanton breaks the symmetry down to $SO(4)$.  The broken generators provide collective coordinates for the instanton, moving the center of the instanton around in the background space-time.  In the AdS case, the measure for these collective coordinates is concentrated at the boundary, which suggests that it is improper to think of the boundary as a place where the metric is frozen and has asymptotic symmetry.   Thus, it is unlikely that a model of an AdS minimum below the great divide has a CFT dual.

Another indication for this is that, with the exception of free $p$-form gauge potentials in $2p + 2$ dimensions, all known CFTs admit relevant perturbations.  This means that any AdS dual of a CFT has a scalar field potential that has a saddle point, with at least one negative direction - a Breitenlohner-Freedman allowed tachyon.  Instantons starting at such a saddle point behave in a very different manner from what we described above.  In AdS/CFT language, both the normalizable and non-normalizable modes fall off at infinity.  In the language of our analog cosmology, the expanding universe is trying to approach a positive c.c. minimum.  It does so independently of the choice of $f(0)$, and for every choice of $f(0)$ the asymptotic solution contains the dominant power law falloff.  The exponential expansion of space in the cosmological analog, is precisely the exponential growth of the volume of spheres in Euclidean AdS.  

The AdS/CFT correspondence tells us that a solution with the leading power law behavior corresponds to a perturbation of the Hamiltonian of the CFT by a relevant operator, rather than a state in the unperturbed theory.   This leads to a bit of a puzzle, since the Lorentzian continuation of the geometry, would seem to be singular.   Since it is not strictly connected to the discussion of Eternal Inflation and the landscape, I will defer the discussion of this puzzle to an appendix.   Suffice it to say that it provides yet another example of the fact that naive extrapolation of instanton physics from quantum field theory to quantum gravity is unjustified.

We now turn to the case where the false minimum has positive c.c., which is the basis for the theory of Eternal Inflation.  Loosely speaking, EI theorists argue that the instanton process is a decay, which occurs locally, anywhere in the exponentially expanding space of de Sitter space-time.  Multiple bubbles can nucleate, but space expands exponentially so that there is always an untouched inflating region.  The global space-time is claimed to be a distribution of causally disconnected bubbles, which approaches a fractal in the infinite future.  Each bubble settles in to a meta-stable minimum of the potential, which might undergo further decays.  The object of theoretical physics in such a system is to estimate the probability that the world looks like our observations.  In doing this, we must ask which of our observations is a prior.  Do we require life?  Do we require intelligence?  Civilization?  Must the life be ``life of our type"?  .     I will make some comments about these issues later on, but first of all I want to investigate the mathematical meaning of the theory.   The practitioners of EI have run into something called  The Measure Problem.  It's fairly obvious that the EI picture predicts an infinite number of each type of bubble.  Various ways of cutting off these infinities and then taking limits, lead to wildly different predictions.

I claim there are more serious problems with the picture.  A profound problem has to do with how one interprets the predictions of quantum mechanics for a system, disjoint parts of which never interact in the far future.  This has been discussed by Bousso and Susskind\cite{qmmultiverse}.  Here I want to concentrate on the technical problem of whether the Coleman De Luccia formalism really leads to the EI picture.

The first surprise, which engenders all of the novelty of the de Sitter case, is that the Euclidean continuation of de Sitter space is the four-sphere, which is compact.  There is no infinite boundary on which we can insist that the instanton approach dS asymptotically.  Let's first consider a transition between two dS spaces, with different values of the c.c. .  Following our analysis above, we see that, starting from $z =0$, with $f_z (0) = 0$, we {\it always} reach a point where $a_z = 0$ and the analog universe starts to re-contract.  To avoid a singularity when $a$ reaches zero again, we use our remaining boundary condition to insist that $f_z$ vanish at the second zero of $a$ as well.  The solution is an ovoid, with a profile of the scalar $f(z)$ which reaches neither dS minimum.   It's easy to show that the action of this instanton is negative, just like that of dS space.  This is not disturbing.  In ordinary quantum field theory one always subtracts the action of the false minimum from that of the instanton, in order to get the negative logarithm of a probability.  In Euclidean functional integral language, this is because the probability is the ratio of two functional integrals.  The quantities
$$ S_I - S_{dS_i},$$ are both positive and can be used to define two probabilities, whose ratio is $$e^{ - \pi M_P^2 (R_F^2 - R_T^2) },$$ where $R_{T,F}$ are the radii of the two dS spaces.

The Lorentzian continuation of the instanton has {\it two} expanding bubbles, which are causally disconnected.  In one bubble, any geodesic observer sees a space-time which rapidly approaches the true dS minimum, while the other approaches the false minimum.  Future infinity in the Penrose diagram is divided into two causally disconnected patches. Traditionally, this has been taken to imply the EI picture.  Both sides of the diagram settle in to a dS minimum, and if one restricts attention to a single observer's causal diamond, the result is a stable equilibrium state.  Traditionally, the part of the Penrose diagram which settles into the false vacuum has been interpreted as the observations of the ``heretics of the false vacuum", observers who stay outside the expanding bubble because dS space expands exponentially.  Brown and Weinberg (BW) have given an alternative interpretation of this diagram, based on a thermal picture of dS to dS tunneling.  They work in the approximation of field theory in a background dS space-time, where everything is mathematically well defined.  It's important to emphasize that the BW picture is (the semi-classical approximation to) an exact and complete description of field theory in dS space, in which the fundamental questions of EI {\it never arise}, at least not in the range of validity of low energy effective field theory.

Brown and Weinberg\cite{BW} have given a clear and coherent explanation of all the properties of the $dS\ \rightarrow dS$ transitions by choosing a potential that is dominated by a large constant, so that there is little change in the geometry.  They argue that in this limit, if all dimensionful parameters are much less than the Planck scale, the tunneling problem reduces to that of quantum field theory in dS space, with fixed c.c..  In QFT it is known that the exact quantum theory, defined by analytic continuation of the Euclidean path integral on the sphere, is Hamiltonian quantum mechanics in the static patch, whose metric is

$$ds^2 = - (1 - \frac{r^2}{R^2}) dt^2 + \frac{dr^2}{ (1 - \frac{r^2}{R^2}) } + r^2 d\Omega_2^2 .$$

The state defined by the Euclidean path integral is thermal, with temperature $T = (2\pi R)^{-1} $.  The surface $r = R$ is null and is called the cosmological horizon.  Introducing tortoise coordinates, it is easy to see that there is a continuum of states near the horizon, which have arbitrarily low energy, as measured by the static Killing vector $\partial_t $.  It's important to recognize that we can view the thermal ensemble of particles at values of $r < R$ as arising from interaction with this bath of states near the horizon.  In QFT the entire Hilbert space of the theory formulated in global coordinates, which appears to contain an infinite number of causally disjoint horizon patches in the far future, is really just two copies of the degrees of freedom in a single horizon, arranged in a thermo-field double state\cite{sussetal}, called the Bunch-Davies state.  Any process taking place in the ever growing spatial volume of the global coordinate slices, is unitarily equivalent to a process describable in terms of the near horizon states and the states in the bulk of a single patch.

Given this picture, Brown and Weinberg explain the features of the CDL instanton in the following manner.  In a thermal system, tunneling is thermally assisted.  We must calculate the probability of going over the barrier starting from any state, and then average over states, with a Boltzmann weight.  The CDL instanton is the leading contribution to this average probability.  It starts a finite distance up the barrier because this is where the compromise between tunneling enhancement and Boltzmann suppression is reached.
There are two probabilities to calculate, because in a thermal ensemble, the lowest (average) energy state in the true minimum can be thermally excited to jump back over the barrier.   The CDL instanton describes both these processes, depending on which subtraction we make.  As we will see, this is a consequence of the principle of detailed balance.  The two Lorentzian continuations describe the classical motion after these two different tunneling events, not different observers' view of the same tunneling event.  The Hawking Moss solution, which just sits on top of the barrier, describes thermal excitation with no tunneling.  For some potentials it is the dominant process of transition, and if the potential maximum is flat enough\footnote{It is possible that potentials this flat do not exist in real theories of quantum gravity.} it is the only mechanism of transition.

All of this is very clear, and there seems no reason to assume that the same rules do not apply to a theory of quantum gravity in dS space.  In particular, I want to stress that {\it within quantum field theory} there is no reason to talk about the region outside the dS horizon.   There is a completely consistent Hamiltonian quantum mechanics of dS space, which deals only with particles within the horizon, and the near horizon continuum states.  Other descriptions of the same physics in global or flat coordinates are simply different time evolution operators (which are time dependent, and do not commute with the static Hamiltonian), operating on the same space of states. This is the quantum version of the familiar statement that the Cauchy problem is well posed within the horizon, as long as we supply boundary conditions on the horizon.  

Another feature of QFT in dS space, which is not stressed by Brown and Weinberg, is that the idea of instanton collective coordinates and the dilute gas approximation breaks down.  A single field theory instanton certainly has a collective coordinate on the four sphere\footnote{This is modified in the quantum theory of gravity.  We will discuss this below.}, but the manifold is compact so the idea of independently variable collective coordinates is no longer valid.  If the size of the instanton is much smaller than the dS radius, then there is an approximate notion of relative collective coordinate.  There can be roughly $(R/R_{Inst})^4$ independent instantons. However, there are forces between instantons which are either repulsive or attractive.  Attractive forces lead to a collapse to a single instanton.  In the example of attractive instanton-anti-instanton forces in tunneling between degenerate vacua in quantum mechanics, the resolution of this collapse requires one to understand the full perturbation series around a single instanton and provide a procedure for resumming it\cite{zinn}.  It's clear that we will never have an analogous prescription for CDL instantons, until we really understand quantum gravity.  If the forces are repulsive then the multi-instantons will form a sort of crystal on the four sphere and there will not be independent collective coordinates.

If we agree to ignore the complications of forces between instantons, then the interpretation of all of this in the single horizon, thermal, picture seems fairly straightforward.  Multi-instantons correspond to independent tunneling events, at different places and times in the horizon.  Bubbles nucleate and grow.  Their centers also move on geodesics.  It's clear that there can't be more than of order $(R/R_{Inst})^3$ visible to the observer at any fixed time, and in a time of order $R$, a typical bubble leaves the observer's horizon.
Thus $(R/R_{Inst})^4$ is roughly the number of bubbles an observer can verify the existence of.  Note however that most of these bubbles have no asymptotic effect on the observer.  Only the bubbles that truly intersect its world line are relevant to the question of how many transitions an observer sees.

So far we have neglected the rate of tunneling in our analysis.  The probability per unit volume per unit time of an instanton transition is really $e^{ - S} R_{Inst}^{-4}$, so it's only when $e^{-S} (R/R_{Inst})^4 > 1$, that the multi-instanton effects can really be interpreted as independent tunneling events, rather than tiny corrections to the single instanton amplitude, which go way beyond the semi-classical approximation and should be neglected.  In typical field theories, with potentials of the form $\mu^4 v (\phi / M)$, we have $R/R_{Inst} = m_P / M$, while the instanton density is truly exponentially small, $e^{-S} < 10^{-40} $.  Thus for $M > 10^8$ GeV, multi-instantons are completely negligible and cannot be discussed consistently in the semi-classical expansion.

It is interesting to analyze how an observer in QFT would interpret the effect of bubbles that nucleate outside of its horizon, or nucleate inside it but move out through it in global coordinates.  Of course, the latter process never literally happens in static coordinates.  Instead, the excitation we are calling a bubble becomes a part of the near horizon continuum modes.  Although these modes have low energy as viewed from the observer's position, any attempt to investigate their structure involves very short distance physics.  {\it In other words, even in QFT, the description of bubbles ``outside the observer's horizon" involves physics at short distance and high energies.}   The fact that the near horizon region has an infinite number of states in a small space-like region, all of which have low energy from the point of view of the geodesic observer in the causal diamond bounded by the horizon, guarantees that, within QFT, the dynamics of the states on the horizon is sensitive to short distance physics.

Bubbles that nucleate outside of a given horizon volume are simply included in the path integral which defines the thermal state for a given observer.  Since all static observers are related by dS invariance there is no observable in conventional QFT, even though it admits the existence of independent degrees of freedom in an infinitely growing collection of horizon volumes, which measures any of the observables that are talked about by the practitioners of EI.
If one actually insists on making those observations, one goes outside the bounds of low energy effective field theory.
In order to talk about a large number of tunneling events we must have $$K =  e^{3 N_e} e^{-S} \gg 1.$$  EI assumes that these tunneling events ``have occurred" and can be thought of as classical evolution in space-time.  Thus, in the quantum state implicitly invoked in EI, one has acted with a large number $> K$ of operators , on the Bunch Davies vacuum.  The number is larger than $K$ because it might take measurement of more than one observable to pin down which event occurred.  For example, in a field space of large dimension, we have to measure the values of each of the fields in order to know which transition has occurred.  

From the point of view of any given observer, these measurements correspond to perturbations of that observer's horizon, and are singular there.  Thus, the state implicitly assumed in EI is NOT the Bunch Davies vacuum. It will lead to singularities in calculations in a single horizon volume, which are sensitive to higher dimension operators in the effective field theory.  

There is not a contradiction between this statement and the claim that the Euclidean functional integral on the sphere is cutoff independent up to the values of the coefficients of relevant and marginal operators in the QFT.  The functional integral on the sphere computes the Lorentzian expectation values in the causal diamond, {\it subject to a particular boundary condition on the horizon.}  This boundary condition corresponds to summing incoherently over the horizon states subject to a condition fixing the average energy measured by the geodesic observer.   It is well known that more general boundary conditions on the horizon lead to singularities there, and physics with such a boundary condition will be sensitive to the detailed short distance
corrections to our effective QFT.   These more general boundary conditions correspond to statements of less complete ignorance of the dynamics of states near the horizon. 

Many authors like to dismiss this sort of observation as a defect of the static coordinate system, pointing to the perfectly smooth global coordinates.  My own suspicion is that the same problem is carefully hidden in global coordinates and is related to the fact that very short wavelengths on the sphere at the throat of dS space, get blown up to large size by the exponential expansion. 
Daniel Harlow\footnote{Private communication} has suggested a calculation in global coordinates, which might shed some light on this picture.  Let us calculate the amplitude, starting with the a false BD vacuum at the dS throat, that at global time $t$ we have $e^{3Ht} e^{-S}$ nucleated bubbles of true vacuum.  In the semi-classical approximation, this requires a, possibly complex, classical solution connecting the two field configurations.  A purely real classical solution does not exist because it would correspond to a backward evolution of multiple critical bubbles into a single horizon volume, and if $t$ is large enough, those bubbles simply don't fit.  This problem becomes worse and worse as $t\rightarrow\infty$.  Thus, I would conjecture that the relevant configuration has a large imaginary part in its action, which increases with $t$.  If this is true, the probability that the Bunch Davies vacuum actually predicts the classical space-time inherent in the EI picture is exponentially suppressed as $t \rightarrow \infty$.
This is the flip side of our previous claim that the classical space-time of EI differs from the BD vacuum by the insertion of a huge number of operators, increasing exponentially with $t$.  We conclude that
 {\it the global description of EI is not compatible with the approximations inherent in effective field theory, and is not predicted by the standard renormalized QFT computations in dS space.}

We've gone about as far as we can with the approximation of QFT in a fixed dS background. The QFT description of CDL tunneling between two dS spaces, gets two related facts wrong, which I believe are clues to the proper theory of dS space.  The first is that the ensemble of near horizon states has infinite entropy below any static energy, while Gibbons and Hawking, by analogy with black hole physics, assert that the entropy is finite and given by $\pi (RM_P)^2$.  This is precisely the same as the analogous paradox for black holes, and we now have pretty overwhelming evidence that the area formula for black hole entropy is correct.

Related to this is a problem with the formula for detailed balance.  The CDL calculation relates the rates for forward and backward transitions by a factor of $e^{ - \Delta S}$, where $\Delta S$ is the entropy difference between the two de Sitter spaces.  This is the appropriate formula for the principle of detailed balance at infinite temperature.  The simplest interpretation of this fact\cite{tbwf} is to say that, consistent with the covariant entropy bound, 
empty dS space is the gravitational configuration representing the maximally uncertain density matrix for a system with a finite dimensional Hilbert space. The CDL formulas are then interpreted as the proper, infinite temperature, form of the principle of detailed balance.

The formula for dS-Schwarzschild black hole metrics gives additional support to this interpretation, and explains the fact that QFT in dS space appears to be a theory at a unique, {\it finite temperature}, $(2\pi R)^{-1}$.  The black hole metric is 

$$ds^2 = - (1 - \frac{R_S}{r} -\frac{r^2}{R^2}) dt^2 + \frac{dr^2}{ (1 - \frac{R_S}{r} -\frac{r^2}{R^2}) } + r^2 d\Omega_2^2 .$$ It has two horizons, determined as the positive roots of a cubic equation with no quadratic term.
Comparing coefficients we find $$R^2 = R_+^2 + R_-^2 + R_+ R_- .$$ This implies that a black hole in dS space (and, by implication, {\it any localized excitation}) has an {\it entropy deficit} relative to the vacuum.  For small $R_-$, we have $R_- \sim R_S$ and $R_+ \sim R$, so the entropy deficit is
$$ \pi R R_S = 2\pi R M .$$ This is precisely the entropy deficit one obtains for a system whose Hamiltonian has eigenvalue $M$, at temperature $(2\pi R)^{-1} $.  

The simplest interpretation of this observation is that a localized excitation of the dS vacuum ensemble corresponds to a freezing of some of the degrees of freedom on the horizon, and that there is a Hamiltonian for the localized degrees of freedom alone, whose eigenvalue degeneracies are correlated with the eigenvalue in the way indicated by the black hole entropy deficit formula.  The maximally uncertain density matrix for the full system, will look like a thermal density matrix at the dS temperature, for localized excitations.
A rather explicit model with precisely these properties \cite{bfm} is reviewed in Appendix B.

Now let's turn to dS decays "to a negative c.c. region of the potential".  First of all, this language is rather misleading, as it was for CDL instantons for flat space below the great divide.  The Lorentzian continuation of the instanton is singular.  The analog universe re-collapses and the field experiences anti-friction, which kicks it out of the basin of attraction of the negative c.c. region.  
In the dS case, in contrast to flat space, this transition always occurs.  The point is that the field does not have to start in one of the minima, so we can use the two initial conditions to avoid a crunch in the Euclidean analog cosmology.  The only way the CDL instanton can fail to exist, is if the potential is too flat at the maximum.  Even in this case we have the Hawking-Moss thermally activated tunneling.

However, we do see a dramatic difference between dS minima which, when shifted down to zero c.c., have a positive energy theorem, and those which don't.  In the latter case, the decay probability is independent of the dS radius when it goes to infinity, while in the former, the probability is of order $e^{ - \pi (RM_P)^2}$, which is what we expect for a finite system, which makes a transition to a low entropy state, analogous to ``all the air collecting in the corner of the room".  This interpretation is reinforced by the observation that for large $R$, the maximal area causal diamond in the collapsing region has an area $\ll 4 \pi R^2$.  Thus, for potentials above the Great Divide, we can assert that the CDL decays to negative c.c. crunches, are not decays at all.
The crunching region is interpreted as a state of low entropy, consistent with the CDL formula for the transition probability into it.

In this interpretation, the reverse process of tunneling from the crunch to dS space, although it is not describable by semi-classical physics, will occur at a rather rapid rate, given by the principle of detailed balance, with the entropy of the crunch computed in terms of the area of its maximal causal diamond.

Finally, I want to comment on the question of collective coordinates for the CDL instantons, once we give up the approximation of QFT in dS space.  Here there is no symmetry argument that they exist, because a compact manifold has no asymptotic symmetries in quantum gravity.  The only argument I know is due to Matt Kleban: if we make both the approximation that the instanton radius is $\ll R$, and the Coleman thin wall approximation, then multi-instantons correspond to a sphere with a number of chordal caps cut off of it .   This shows that the number of instantons is again bounded by $(R/R_I)^4$, and that the forces between instantons fall off exponentially, as was the case for field theory on the sphere.   The same issues with integrating over the collective coordinates arise: it never seems to be justified.  In the thermal, causal patch interpretation, only instantons which collide with the causal diamond of the observer, may be studied in low energy effective field theory.  In a true theory of quantum gravity the QFT observation that all physics {\it may} be understood within a causal patch is replaced by The Holographic Principle:  physics outside the causal patch is just a gauge copy of physics inside.  Thus, the Holographic Principle is a strengthening of the remarks we made about QFT in dS space.  From the QFT point of view, effective field theory breaks down when we consider too many instantons in the global coordinates of dS space\footnote{The quantitative measure of ``too many" depends on our choice of cutoff.  The measurement of a certain number of instanton transitions outside the horizon corresponds to a measurement of a certain density of local operators in the near horizon region of the static patch.  A spatial UV cutoff of the static patch field theory replaces the horizon by a stretched horizon and puts a length cutoff on that surface.  When the density of local operators is of order the cutoff density, effective field theory has broken down.  All of the infinities of EI in global coordinates (within the approximation of a fixed metric) are related to singular UV boundary conditions on the static patch horizon.}. In QG, the space outside the horizon has no gauge invariant meaning.

To summarize: the definition of field theory in dS space, even without invoking quantum gravity, suggests that the only description of EI which is amenable to honest low energy field theory analysis  is one which concentrates only on CDL transitions that occur within, or perhaps on the boundary of, the causal diamond of a single geodesic observer.  Quantum gravity, in the form of the statement that the dS entropy actually represents the log of dimension of the quantum mechanical Hilbert space describing the system, resolves various puzzles of the field theoretic picture, while reinforcing the idea that the proper theory of EI should describe only the maximal causal diamond of a single observer.  These observations suggest an interpretation of empty dS space as the maximally uncertain density matrix on a finite dimensional Hilbert space, whose dimension is the exponential of the dS entropy.  Localized excitations are low entropy states of the system, as shown by the Schwarzschild dS black hole entropy formula.  The Hamiltonian of a geodesic observer, which is described in the appendix, has the form
$$H = P_0 + \frac{1}{R} V ,$$ where $V$ has a bound of order $1$.  $P_0$ measures the energy of localized particles and black holes in a single horizon volume.  The bound on $P_0$ is the Nariai black hole mass, of order $R M_P^2$.  For eigenvalues well below this bound, the entropy deficit of the eigenspace with energy $P_0 = E$ is $2\pi R E$, so that the dS vacuum is a thermal state for $P_0$, at the dS temperature.  This explains why the dS temperature is unique.

Our survey of CDL instantons has led to the following conclusions.  For minima of the potential which are below the Great Divide, CDL instantons are genuine instabilities with a number of puzzling properties.  The final state of a CDL decay depends on the initial state and on events which are taking place far from the site of bubble nucleation.  Bubbles can be swallowed by black holes, and the interiors of black holes are unaffected by bubbles nucleated outside of them.  Bubbles nucleated in, or swallowed by black holes do not grow.  The asymptotic future of space-time consists of a set of causally disconnected singular regions whose number and nature depends on the initial state.   AdS minima below the great divide do not correspond to CFTs.  The boundary of such space-times is dominated by instantons.

In Ads/CFT, the AdS point is always a saddle point, because the CFT has relevant perturbations.  In this case, as described in the Appendix, CDL instantons are not decays of the initial system, but correspond to perturbations of the original theory by a relevant operator that is unbounded from below.  The instanton has a free parameter, corresponding to the strength of the relevant coupling. Depending on the choice of boundary conditions, the resulting theory {\it might} make sense if the QFT is restricted to Euclidean or Lorentzian dS space, and the relevant coupling is not too large.

For positive c.c. , CDL transitions always exist, at least the pure thermal activation represented by the Hawking Moss instanton.  Below the great divide we have the same problems that exist for all values of the c.c. .  Above the great divide, all transition probabilities are consistent with the idea that the lowest lying dS space is a system with a finite number of states, and the transitions are all reversible and short lived excursions of this equilibrium state into low entropy catastrophes.

\section{The Phenomenology of Eternal Inflation}

There are, in my opinion, two primary reasons that the global view of eternal inflation has dominated most of the literature on the subject.  The first is that the phenomenological models of inflation, which have had great success in explaining the properties of the cosmic microwave background (CMB) and large scale structure in the universe definitely require one to treat independent horizon volumes of an exponentially expanding universe as independent quantum systems.  We measure the relative amplitudes of fluctuations in different inflationary horizons, when we look at the CMB.
The holographic view of dS space as a single horizon volume does not seem justified.

It isn't, but this has been understood since Bousso's first paper on the subject.  In slow roll inflationary models, or any model where classical evolution passes through an approximate de Sitter phase, which is succeeded by sub-luminal expansion, late time observers have causal diamonds whose size is bounded by the final value of the c.c., rather than the one controlling the temporary inflationary phase.  {\it de Sitter space is not inflation, and inflation is not de Sitter space.}.   It does not, therefore, follow that the phenomenological success of inflationary models requires us to accept the global view of EI.

The second psychological reason for the prevalence of the global view is the success of Weinberg's prediction\cite{weinberg} for the value of the c.c. based on anthropic reasoning.  Linde and I were the first to suggest an anthropic explanation for the value of the c.c. based on inflationary models\cite{tblindecc}, but within the context of the string landscape, or most any contemporary view of global EI, I don't think anthropic reasoning leads to good phenomenology.

There are two logically independent reasons for this.  The first, and most talked about, is the measure problem of global EI.  The slogan that ``everything that can happen, will happen, an infinite number of times" leads to a confusing question of how to compare infinities in determining the predictions of EI.  Most naive guesses, based on the global picture, lead to immediate conflicts with the crudest measures of experimental data.  Most observers are predicted to be evanescent thermal fluctuations, with memories of a past that never existed.  Or, using a different measure, one predicts that most post inflationary universes have just been formed, so that an old universe like our own is super-exponentially\footnote{Super-exponential refers to numbers of the form $e^{\pm (e^{N})}$, with $N$ an integer of moderate to large size.} improbable.  These questions have been dealt with in the EI literature, and I will make brief comments about them below.

A more serious issue\cite{bdg} is the extent to which anthropic reasoning can predict more than  two or three things about the model that describes our universe. If it can't, then a landscape model in which more than two or three low energy parameters vary over the landscape, is in violent disagreement with experiment.  The point is that the standard model, including operators up to dimension $6$, has a huge number of parameters, many of which could be orders of magnitude larger than they are without disturbing the parts of the universe that are necessary for our existence.

Anthropic arguments are of two types.  They can be very general.  A universe which goes from Big Bang to Big Crunch or exponential expansion on short time scales, is uninhabitable by any conceivable creature.  Weinberg's argument is almost of this type, but does require knowledge of the primordial dark matter density and fluctuation spectrum.  It also requires a modest assumption that one cannot have life without galaxies.  Note that even here we have a problem, because the bound is much more sensitive to the size of primordial fluctuations than it is to the c.c. .    

The second type of anthropic argument requires us to specify that we are looking only for life, which obeys the same kind of chemistry as ourselves.  This is a drastic and somewhat arbitrary cut on the vast array of possibilities presented by the string landscape, but it's certainly the best we can ever hope to do within this framework.  It goes beyond hubris to imagine that we will ever be able to solve for the details of low energy physics and biology for a general product gauge group and set of matter representations, with not a scintilla of experimental data.

It's important to realize that these anthropic arguments take on different forms at different levels of low energy effective field theory.  Arguments about life of our type depend on chemistry, and on some of the properties of stars and galaxies.  The latter require nuclear physics at the MeV scale, and, for supernovae, some details of neutrino physics.   At one's most optimistic one might argue that one needs so many coincidences and fine tunings of nuclear levels that one is forced to the $SU(3)\times U(1)$ theory of the strong and electromagnetic interactions, with some constraints on the masses of the up and down quarks.  The electron to quark mass ratios also have a potential anthropic explanation.  The weak interactions come in at best in the form of $4$ fermion couplings, with only the ratios of couplings to gauge boson masses determined by anthropic considerations\footnote{And one should not forget the possibility of an anthropically allowed {\it weakless} universe\cite{weakless}.}.  They do not have to be unified with electromagnetism and they do not have to be chiral, because anthropic considerations will require the up and down quark masses to be small compared to the Planck scale.  Couplings that lead to proton decay, and lepton number violation can be many orders of magnitude larger than they are in the real world, as can strong CP violation.  Neutrino masses could also be much larger than they are\footnote{No anthropic argument of which I am aware requires more than one flavor of neutrino. Furthermore, primordial nucleosynthesis of precisely the sort indicated by observation is {\it not} crucial to life.  We need to preserve only enough properties of stars to lead to stellar synthesis of the heavy elements, and to 
at least one star like our own sun.  In typical landscape models the ratios of universes of different types are super-exponential numbers, so we really want the minimum of anthropic constraints.  If models with the number of Sol type stars require more tuning of parameters than those with only one, this will far outweigh the unlikeliness of developing life with only one shot at it in a given universe.}.  It goes without saying that an honestly anthropic evaluation of the landscape would not predict Grand unification of couplings.  

Flavor becomes much more of a puzzle in this framework than it already is.  Indeed, if we assume an anthropic explanation for the up and down quark masses, then all other generations are predicted to be at the Planck scale.
Scott Thomas has suggested the possibility that muons penetrating the atmosphere might have been crucial to the mutation rates that drive evolution.  Even if this is the case, in a landscape driven only by anthropic considerations, it is much more likely that we would get a muon by tuning the mass of a single vector-like fermion than by introducing a whole new fermion generation.  Thus, the landscape makes Rabi's famous question about who ordered the muon, even more poignant.

Thus, it would seem that a purely anthropic approach to the landscape, without prejudice induced by the benefit of experimental data, predicts a low energy theory at the $1-100$ GeV scale, which is radically different from what we observe.  One is forced to fall back on claiming that the String Landscape is special and on trying to preserve the role that symmetry plays in more standard approaches to particle physics.  However, the evidence is that symmetry is very rare in the string landscape\cite{dinesun}, so one is forced to claim that this evidence is somehow not representative of the true nature of the landscape.

The issue of supersymmetry is similar.  The MSSM adds over 100 new parameters to the standard model at the level of dimension $4$ operators, and many more if we look at dimensions up to $6$.  Most of these operators are very highly constrained by experiment but not very constrained by anthropic considerations.  In other words, SUSY at any scale below the Planck scale (or perhaps the unification scale) makes the phenomenological problems of the landscape picture much much worse.

The initial reaction to these facts was an attempt to argue \cite{douglassusskind} that low energy SUSY was very unlikely in the landscape.
Dine argued by contrast that low energy SUSY was likely to be the most probable way to explain the value of the c.c. and the electroweak scale, within the landscape context.   At least one of the authors of \cite{douglassusskind} has come around to the point of view that low energy SUSY is in fact likely\cite{douglas}.

There has not been much discussion of these phenomenological problems of the landscape.  I believe the reason is that those working on the landscape program consider it to be part of the measure problem.  That is, until one has come to some conclusion about how to calculate probabilities in EI, it is premature to base phenomenological conclusions on mere counting arguments in the string landscape.  I will go on to a quick survey of the measure problem in the next section, but want to emphasize that in order to resolve these phenomenological problems, a solution of the measure problem that does not lead to violent contradiction with experiment, must find principles to replace the symmetry considerations of non-EI physics.  Indeed, with the exception of the c.c., we have plausible symmetry arguments that explain all of the apparent fine tuning in the supersymmetric standard model\footnote{Model builders are worried because recent LHC data has ruled out the simplest supersymmetric models, in the sense that one or two parameters require a fine tuning of about $1 \%$ (some models are a few orders of magnitude worse than this) in order to explain the data.}.  In many cases there is no agreed upon explanation, only because so many different models can be built which explain the tunings.  By contrast, in the absence of new principles, landscape models require fine tuning of many orders of magnitude, in many different parameters, in order to explain the data of our world.  The phenomenological triumph of explaining the value of the c.c. has come at the cost of a huge set of new puzzles, which are avoided by non-landscape physics.

\section{Theories of the Landscape}

This section will be brief, and the reader who really wants to understand the details should consult the extensive literature\cite{landscapelit}.  In addition, I will concentrate on ideas, which focus attention on measurements made within a single causal diamond.  Quite apart from the remarks I made above about the proper interpretation of CDL tunneling from dS space, Bousso and Susskind\cite{qmmultiverse} have pointed out a conceptual problem with the conventional global EI picture.   Tunneling is a quantum phenomenon, but the conventional EI picture is that of a classical space-time, where all of the tunneling events have actually happened.  We usually justify such an interpretation of quantum wave functions by appealing to decoherence.  Some set of quantum events actually happen because they affect the behavior of macro-systems, for which quantum interference is measurable only over recurrence times that are super-exponentially larger than the age of the universe.   However, in EI we are insisting on talking about times that are infinitely long and all of this must be rethought.   Where in the EI multiverse is the effective classical observer, which can decohere all the CDL tunnelings into an effective classical space-time?

The only proposal for answering this question, which seems sensible to me, is the one dubbed The Census Taker's Hat\cite{census}.  Here one concentrates on a particular causal diamond (the {\it census taker's diamond}) in the hypothetical EI multiverse, and notes that, in addition to the CDL transitions experienced by a particular time-like world line in this causal diamond, the EI picture leads us to expect evidence for the multiverse to 
appear through collisions of other bubbles with the census taker's final bubble.  In order to have a universal answer to questions of counting one assumes that the final bubble along world lines in this diamond has zero c.c.. All the evidence in string theory, as well as field theoretic self consistency of the vanishing c.c., suggests that locally the interior of this final bubble must resemble one of the super-Poincare invariant space-times, which are known solutions of string theory\footnote{In fact the class of such solutions is large and the space of solutions is not known to be connected, and is definitely not a manifold.  There are many perturbative heterotic string constructions based on supersymmetric asymmetric orbifolds which have never been fit into any sort of string duality picture. Some of them have few moduli in perturbation theory.  The weakly coupled region of the moduli space of such theories is just as good a candidate for the local physics of a Census Taker's hat as the ten and eleven dimensional backgrounds that are tacitly assumed in the literature.  In other words, even if the Census Taker's Hat turns out to be a good theory of Eternal Inflation, it may not be unique.  We could find that there are as many such independent models as there are $AdS_4 $ models.}

The time-like trajectories in the Census Taker's diamond have an asymptotically common holographic screen, which is the two sphere at spatial infinity of  the last CDL bubble they experience.  This screen has infinite area, so the quantum theory has an infinite dimensional Hilbert space, and can plausibly contain precise observables, which serve to decohere the quantum processes of CDL nucleation in bubbles that have collided with the causal past of the screen.  The measure implied for EI is the causal patch measure of Bousso\cite{causalpatchmeasure}.   Remarkably, this measure is equivalent to a global measure called light cone time\cite{lightconetime}.  This is an example of what Bousso calls {\it global-local duality}, in which observations of bubble collisions in a single diamond, can reproduce features of the global picture. 
It is the basis on which the global picture of EI might eventually be justified.

I don't have space to go into all of the details of this picture, which are at any rate changing rapidly, but I want to comment on the current ideas about how physical predictions are supposed to come out of this framework.  The basic equation that essentially all approaches to EI end up with has a Markovian form

$$P_{n+1} = P_n + \sum \gamma_{nm} P_m - \sum \gamma_{mn} P_n .$$
Here $\gamma_{nm}$ is the rate of transitions from minimum $m$ to minimum $n$ of the potential.  These are related by detailed balance $$\gamma_{nm} = \gamma_{mn} e^{S_n - S_m} ,$$ where $S_n$ are entropies of different states, if the transitions are above the Great Divide.

Terminal states are defined to be those labelled $T$ for which the transition probabilities $\gamma_{mT}$ are all zero.  Many authors include negative c.c. crunch states in this category.  I believe those for which the $\gamma_{Tn}$ transitions are above the great divide have finite entropy and do not belong in this category, while those below the great divide may not make sense at all.   One example of a terminal state $T$ is the infinite entropy census taker's hat.   Actually (see the previous footnote) there are likely to be multiple inequivalent census takers.   The simplest assumption in that case is that there a multiple inequivalent landscapes.  We know so little about this subject\footnote{Indeed, I've spent most of this note trying to convince the reader that the landscape doesn't exist.  In case it is not obvious, we are suspending that disbelief for the course of this section.} that I won't try to investigate more complicated scenarios. 

Susskind has recently argued\cite{arrow} that if there are no terminal states,
then the long time behavior is that of an equilibrium system.  If the total number of minima is finite then any such equilibrium has the problem that the most common observers, according to the equilibrium probability distribution, are thermal fluctuations with false memory traces.  The landscaper's fascination with the anthropic principle again leads to disastrous predictions.

The census taker's hat resolves this difficulty, but only with rather drastic assumptions.  First one must assume that no observers can exist in the census taker's hat itself.  The hat is a cosmological space-time with infinite spatial sections on which the energy density is non-zero.   It seems to me that if one is willing to worry about thermally fluctuated observer's with false memories, then it is likely that one must worry about any Turing machine, which can simulate such an observer.   The hypothesis of artificial intelligence says that such a machine, running a program of sufficient complexity, will feel self aware.  We don't know if this is correct, but if we are willing to indulge in fantastic speculations like Boltzmann Brains, it seems incorrect to deny a plausible speculation in another domain of science.

Local physics in the census taker's hat is asymptotically supersymmetric, and this is probably enough to rule out the existence of physical observers ``of our type".  However it's far from clear that no complex Turing machines can exist and recur into the indefinite future.

We get another try at falsifying this model when we look at the most likely pre-hat state of the multiverse.  This is the {\it dominant eigenvector} of the transfer matrix of the Markov process.   Since many of the transition  probabilities $\gamma_{mn}$ are super-exponentially small, with super-exponential ratios, the dominant eigenvector is mostly in the direction of the vacuum with the minimum width  $$\Gamma_n = \sum_m \gamma_{mn} ,$$ where the sum includes the terminal hat.  It is likely that this is a nearly supersymmetric dS space, with a very small value of the c.c., which is not our world.   It's again important that this state, which we call the dominant dS state, does not support any complex Turing machines, which could simulate our memories.

Finally, by doing eigenvalue perturbation theory, the world with the dominant observers will be one that is connected to the dominant dS state by a small number of transitions.  Perhaps, in the fullness of time, we could show that it resembled the world we observe.  But I wouldn't hold your breath...

\section{Conclusions}

The notion of quantum effective potential on which the String Landscape is based, has no basis in perturbative string theory, the AdS/CFT correspondence, or Matrix Theory.  All of these point instead to a paradigm where different space-times correspond to different Hamiltonians\footnote{The semi-classical solution of the Wheeler-DeWitt equation, and the general notion that Hamiltonians in General Relativity are defined at spatial or null infinity, are further hints that this is the correct paradigm.}  The Fischler-Susskind mechanism does not lead to a systematic world sheet expansion for even the tamest example of shifted vacua - that provided by pseudo-anomalous $U(1)$ gauge symmetries, where the shifted vacuum is super-Poincare invariant and has a variable small parameter, according to the effective field theory analysis of \cite{dsww}.   In the next most plausible
set of models where the classical supergravity Lagrangian does not capture the correct asymptotic values of space-time fields, AdS solutions of F-theory modified by a ``non-perturbative" superpotential, the Fischler-Susskind analysis fails more dramatically.  The SUGRA solution has a moduli space of SUSY violating Poincare invariant solutions of the ``no-scale" type.  Fischler-Susskind will capture the perturbative corrections to the Kahler potential, which ruin the no-scale backgrounds.  The world-sheet analysis will find a cosmological solution, violating SUSY, for which perturbation theory breaks down at either early or late times, so that scattering amplitudes do not exist.

When the value of the tree level superpotential is small, an effective field theory analysis suggests that a perturbatively invisible exponential correction to the dependence of the superpotential on the Kahler moduli of the F-theory
manifold can stabilize the system and give rise to a supersymmetric AdS compactification. The ratio of the AdS radius to that of the compact manifold can be tuned to be large by exploiting the large multiplicity of fluxes on a topologically complicated internal manifold.   This is the ``string-landscape", but it should clearly be called an ``effective SUGRA landscape", since it has no direct connection to string theory.  A connection might be found by using the AdS/CFT correspondence, but aside from the work of \cite{eva}, this has not been done.  The success of such a program would imply the exact opposite of a landscape picture.  Each ``vacuum state" would be a separate CFT, with a different high energy spectrum, OPE coefficients, {\it etc.}.

There has been a lot of discussion in the literature of ``uplifting" these AdS solutions to meta-stable dS solutions.  There are many technical quibbles one can make about these discussions, but the most incisive criticism of them is that the theory, such as it is, of meta-stable dS vacua, has nothing to do with CFTs dual to AdS space.  Thus, the claim that one can establish the existence of the vast landscape of meta-stable dS vacua, as ``small perturbations" of a theory of a single AdS space, seems manifestly false.  The AdS/CFT correspondence is fairly well understood, and in that context it seems absolutely clear that different solutions of the same low energy effective Lagrangian are different quantum theories.   

The standard theory of EI is mostly based on CDL tunneling, and in that context too it is clear that AdS space is not part of the picture.  When a meta-stable state tunnels into a basin of attraction of a negative c.c. minimum, the classical evolution is a Big Crunch space-time, in which the scalar field does not even remain close to that minimum.  Analysis of such processes leads to a co-dimension one Great Divide in the space of potentials.  In the case of tunneling from minima with non-positive c.c., this Divide corresponds to the existence (or not) of a positive energy theorem for the Minkowski or AdS space.  For positive c.c. there is {\it always} a transition, but we have argued that above The Great Divide it corresponds to a temporary sojourn of the lowest dS minimum in a very low entropy state.  The lowest dS minimum corresponds to the maximally uncertain density matrix of a system with a finite number of states.  I would claim that even in the case where the CDL transition goes to a negative c.c. crunch, there will be rather rapid return to the dS vacuum, with rate given by the principle of detailed balance, using the area of the maximal causal diamond in the crunching region as a calculation of the entropy of that state.

In our discussion of CDL tunneling I emphasized an important fact, which has not previously been noticed. Even in the approximation of quantum field theory in a fixed dS space, the conventional approach to Eternal Inflation is not justified within effective field theory.  The system can support a large but finite number, $e^{-S} (R/L_I)^4$ of instantons\footnote{In making this statement I am neglecting problems with forces between instantons, which require a very delicate non-perturbative discussion even in ordinary quantum mechanics.} when the instanton size is much less than the dS radius.  This has a simple explanation in a thermal picture of CDL tunneling in a single horizon volume.  Instanton calculations involving only these instantons can be studied in low energy effective field theory because the Euclidean functional integral on the four sphere is renormalized by the same counterterms that make the theory in flat space finite.   

However, measurement of a large number of instantons outside of a single horizon volume, corresponds to inserting singular boundary conditions on the horizon of a given observer, and these are sensitive to short distance corrections to the effective Lagrangian.   The thermal interpretation of tunneling in a single horizon volume, recently re-analyzed in a lucid and compelling paper by Brown and Weinberg\cite{BW}, gives a simple explanation of all the peculiarities of this problem and shows none of the characteristic infinities of EI, as long as we stick to the time independent Hamiltonian continuation of the well defined Euclidean functional integral.

The Holographic Principle reinforces and improves on this QFT picture, and leads to a suggestion that the theory of stable dS space is a quantum theory with a finite number of states.  CDL transitions out of the lowest positive c.c. minimum of the potential, when the potential is above The Great Divide, correspond to temporary sojourns in low entropy states, even when the low entropy state is a singular negative c.c. crunch, and the reverse transition does not have a semi-classical description.  

Ignoring these attacks on the fundamental basis of the EI picture, I briefly discussed theories of eternal inflation, with emphasis on the set of ideas developed by Bousso, Susskind and collaborators, who emphasize models based on the observations in a single causal diamond.  Some of these authors emphasize the necessity for eternal inflation to end with a decay into a locally supersymmetric space-time, with vanishing c.c., infinite entropy, and negatively curved spatial segments.  Counting observers in such a model is related to collisions of arbitrary CDL bubbles in the landscape with the bubble in the {\it Census Taker's Hat}.  Crude equations based on this picture lead to a Markov process whose final state is always the hat.  The penultimate state in the process is the dominant eigenvector of the Markov equation, whose decay rate into the hat is superexponentially slower than any other state.  This eigenvector is predominantly a single, almost supersymmetric vacuum, with very small c.c. (much smaller than our own).  Standard eigenvalue perturbation theory for this equation mixes in other minima of the potential, which are obtained from the dominant vacuum by a few, relatively large actions of the hopping matrix.  It is hoped that the most probable minimum, which contains life of our type, in the dominant eigenvector, makes correct predictions about the physics we observe.

We've seen that this is a tall order for a purely anthropic theory, where most of low energy physics can fluctuate from minimum to minimum, will have a very hard time reproducing the Lagrangian of the standard model, especially if there is low energy SUSY.  A wide variety of gauge theories, with no detailed resemblance to the weak interaction part of the standard model can reproduce the nuclear and atomic physics that are necessary for life as we know it.  In particular, life would be fine with a low energy $SU(3) \times U(1)$ theory with vector-like weak interactions and fine tunings of up and down quark and electron mass, no extra generations, and only the ratio of coupling squared divided by vector boson mass squared for the weak sector determined.  The fermion masses are not related to the spontaneous breakdown of the weak gauge group, and electromagnetism is not unified with weak interactions.   It may be that one can do without weak interactions at all\cite{weakless}.   Many lepton and baryon number violating couplings are allowed, with coefficients larger than those indicated by experiment by many orders of magnitude.  Purely anthropic arguments, with a uniform measure over the putative space of at least $10^{500}$ string vacua, are strongly ruled out by experiment.

As far as I can understand, the hopes for avoiding this argument (as opposed to ignoring it) are, at the moment, based on the rate equations determining the distribution of vacua as seen at late times by an observer in the Census Taker's Hat. As noted above, these are dominated by a single eigenvalue of the Markov transfer matrix governing CDL decays in the landscape.  That eigenvalue has a probability close to one of being a highly supersymmetric vacuum with c.c. much smaller than our own, whose total rate of CDL decay is super-exponentially smaller than that of all other vacua.
The small probabilities for other vacua are dominated by the first few terms in hopping matrix perturbation theory around this state, because the matrix elements are all small and wildly different from each other.  Our own world is determined as the dominant term in this perturbation theory, which allows for low energy physics compatible with our form of life.  One hopes that this perturbation theory takes in a small enough part of the landscape that this criterion is predictive, and that the predictions agree with our observations.  One would then describe all the apparent fine tunings of the previous paragraphs as accidental properties of this special state.

It is further assumed that both the Census Taker's Hat, and the dominant, almost supersymmetric, vacuum, which decays into it cannot support large number of recurrences of conscious observers with false memories matching our own.  Supersymmetry of both these states precludes the existence of atoms and nuclei, but one must, I contend, beware of Boltzmann's Turing machines.  Turing's theorem shows that a very simple system can simulate any computer program, if it has sufficiently many registers.  A widely believed conjecture in the artificial intelligence community is that a sufficiently complex program becomes self-aware, and that our own consciousness is simply such a program running on the particular biological hardware that evolution constructed.  Thus, if highly supersymmetric states can support {\it any} complex structures with a number of states equal to that of a thermally fluctuated observer, then there will be conscious observers in both the dominant vacuum, and the hat, which vastly outnumber the ordinary observers, like ourselves, who have good reason to believe that they obey the laws of physics and chemistry that we know.

\vfill\eject

\begin{center}
{\bf Acknowledgements}
\end{center}

I would like to thank W. Fischler, L.Susskind, S.Shenker, R.Bousso, G.Horowitz, E.Rabinovici,D.Harlow, and J.Maldacena, for numerous conversations, which led me to an understanding of the issues discussed in this paper.  This work was supported in part by the Department of Energy.

\section{Appendix A: Coleman Deluccia Tunneling from an AdS Saddle Point}

$AdS_d$ spaces which have a CFT dual admitting a relevant perturbation, all have saddle points in the space of scalar fields, corresponding to Breitenlohner-Freedman allowed tachyons.  The operator dimension/scalar mass map is
$$ \lambda_{\pm} = \frac{1}{2} [(d - 1) \pm \sqrt{(d - 1)^2 + 4 m^2 R^2} ],$$ corresponding to wave functions that fall like
$ e^{- \lambda_{\pm} z} $, in the coordinate system where the Euclidean AdS metric behaves like
$$ds^2 = dz^2 + e^{2z} d\Omega_{d-1}^2 $$ at infinity.  The CFT dual lives in dimension $d - 1$, which form a sphere in these coordinates, and the operator dimension is $\Delta_m = d - 1 - \lambda_- $.  The larger root, with more rapid falloff, corresponds to the expectation value of the operator in the presence of the source.  Perturbations of the theory with the dominant $\lambda_-$ behavior correspond to deformations of the Lagrangian by the operator with dimension $\Delta_m$.  

In the analog cosmology, which solves the same equations as the instanton, we have a non-singular initial condition, which evolves to a dS minimum (but with negative spatial curvature).  The solution is non-singular and settles down to the dS minimum for {\it any} choice of the initial condition $f(0)$ at the point where $f_z (0)$ and $a(0)$ vanish.  This corresponds to the fact that both solutions of the linearized equations fall off at infinity.

The standard AdS/CFT interpretation of this instanton solution is therefore a perturbation of the original boundary theory by a relevant operator.  $f(0)$ parametrizes the strength of that coupling, in units of the AdS radius.
This interpretation of the instanton first appeared in the work of Hertog and Horowitz\cite{HH} .  These authors noted that the instanton satisfied a boundary condition relating the coefficients of the two linearized solutions.  According to the work of \cite{berkoozwitten}, such a boundary condition corresponds to a perturbation by a multi-trace operator.  In the particular case studied in \cite{HH}, the CFT was the IR fixed point of $2+1$ dimensional maximally supersymmetric Yang Mills (SYM) theory, which is dual to 
$AdS_4 \times S^7$.  The relevant operator is the IR limit of $ O = {\rm Tr} \ (\Phi_1^2 - \Phi_2^2) ,$ where $\Phi_i$ are two of the seven scalars in the SYM theory.  The HH boundary condition corresponds to inserting the cube of this operator, which is marginal but unbounded from below, into the Lagrangian.

Maldacena\cite{maldahh} pointed out that the instanton {\it also} satisfies the simple Dirichlet boundary condition, which corresponds to perturbing by $O$ itself.  The difference, according to Maldacena, lies in the boundary conditions that one imposes on {\it fluctuations around the instanton}, when doing the Euclidean functional integral in the bulk.  Similar remarks have been made in work of Barbon and Rabinovici\cite{barbrab} and Harlow and Susskind\cite{harlsuss}.

All of this analysis should give pause to anyone familiar with the cavalier way in which instanton transitions are treated in discussions of EI.  There one typically imagines one is dealing with a situation in which "the non-gravitational QFT discussion of instantons is a good approximation", and takes the existence of an instanton with one negative mode as evidence for instability of a particular state in a hypothetical theory of the Landscape.  Here we see that an instanton corresponds to a change of the theory (the original theory studied by HH is maximally supersymmetric and certainly stable), and that to determine what that change is, one must make a decision about how the fluctuations around the instanton are quantized !

For a Euclidean conformal field theory, flat space, the $d - 1$ sphere and $R \times S^{d-2}$ are closely related manifolds, since they are all conformally flat.  Green functions on one manifold are closely related to those on another, the partition functions are related by the conformal
anomaly.  Only subtle questions, like whether there is a unitary representation of the full conformal group, depend on which of these manifolds we use.  However, once we perturb by an operator that violates conformal invariance, this equivalence is no longer valid.  

A familiar example is Holographic Renormalization Group flow.  When a conformal field theory with a large radius AdS dual is perturbed by a relevant operator on flat space, the result is sometimes related to a static domain wall space-time, connecting two AdS spaces.  This is a flow between two CFTs, each of which has a large radius dual.  On the other hand, if we put the same quantum field theory on $S^{d - 1}$ or $R \times S^{d - 2}$, the RG flow is stopped by the IR cutoff and we flow instead to a quantum mechanics of zero modes.   There is no corresponding domain wall space-time.  

For the HH perturbations, the meaning of the instanton seems to depend both on the manifold on which we compactify, and on the boundary conditions we choose for fluctuations.  The CDL coordinates for the instanton suggest that we should consider the CFT dual on the Euclidean $d - 1$ sphere (where $d=4$ for the HH solution), whose Lorentzian continuation is $d - 1$ dimensional de Sitter space.  References \cite{maldahh}\cite{
barbrab}\cite{harlsuss} point out an interesting interpretation of this choice.   The Lorentzian continuation of the HH instanton is a singular Big Crunch.  If we choose a de Sitter slicing of $AdS_4$ then there is a family of dS slices which asymptote to the light cone that bounds the singularity.  These slices hit the AdS boundary in finite global time, and the coordinate change between the global coordinates and these dS coordinates is singular at the boundary.  

The boundary field theory in dS space has a positive quadratic term in the scalar fields $\Phi_i$ coming from the curvature.  Thus, the origin of field space appears to be a meta-stable minimum of the boundary theory\footnote{We are making an assumption that we can analyze meta-stability in SYM theory, where the scalars are approximately free fields, and that this analysis remains valid when we flow to the IR fixed point, which is dual to $AdS_4 \times S^7$.  }, with both choices of boundary conditions, as long as the coefficient of the relevant perturbation with Maldacena boundary conditions is small enough.   In the HH interpretation of the instanton, this meta-stability is not an ultimate help: the boundary theory has a Hamiltonian which is unbounded from below.  Its functional integral on $S^4$ is not well defined.  For the Maldacena boundary condition, with small enough coefficient, it seems that the theory {\it is} well defined, and it is claimed to provide a non-singular dual to a Big Crunch singularity.

This is an incredibly interesting assertion, and it deserves much more attention than it has gotten.  If it's correct it means that field theory in dS space can contain the physics of certain kinds of space-like singularities.  The question is: which field theories, and how do we tell {\it a priori} whether a given space-like singularity has such a description.  It's clear from the example that it depends on rather subtle questions from the bulk point of view.  Quantizing the fluctuations around the same space-time background in two different ways, seems to correspond to the difference between a background field theory that is well defined non-perturbatively, and one which isn't.  One should also note that, from the boundary field theory point of view, it seems that the theory with Maldacena's boundary condition only makes sense if the coefficient of the perturbation is not too large.  This would seem to indicate some sort of a singular phase transition, as a function of the free parameter $f(0)$ of the instanton.  Such a transition seems completely mysterious from the point of view of the bulk theory.

\section{Appendix B: The Quantum Theory of Stable de Sitter Space}

In the text we demonstrated the following proposition:  Consider a potential with only positive and negative minima, which is above the Great Divide when we add a negative constant so that the lowest dS minimum has zero energy.
Then all CDL tunneling probabilities {\it out} of this state are of order $e^{- \pi (RM_P)^2}$ when $R M_P \rightarrow\infty$.  CDL transitions are then interpreted as low entropy transitions in a system with a finite number of states, at infinite temperature.  The empty dS vacuum is the maximally uncertain density matrix.  The black hole entropy formulas indicate that we should consider the finite dS temperature to be a property of a Hamiltonian for localized states only.  These are low entropy excitations of the dS vacuum with a correlation between the Poincare energy $P_0$ of the state and the probability of finding it in the infinite temperature ensemble which is given by the Boltzmann factor
$$e^{ - 2\pi R P_0}.$$

Here is an explicit model, which gives exactly that structure.
Consider the operator algebra

$$[\psi_i^A (P) , (\psi^{\dagger})_B^j (Q) ]_+ = Z_{PQ} \delta_i^j \delta_B^A ,$$ combined with commutation relations between $Z$ and $\psi$, which make this into a finite dimensional super-algebra, with a finite dimensional representation generated by the fermionic generators. $i$ runs from $1$ to $N$, $A$ from $1$ to $N + 1$ and $P$ runs over $0$ to $K$.  $Z_{00} = 1$.   The fermionic generators are associated with the spinor bundle over $S^2 \times {\cal K}$ with separate cutoffs on the spectra of the Dirac operators on these two compact Euclidean manifolds.

If we consider the $N\rightarrow \infty$ limit we can extract generators of the single particle SUSY algebra by a procedure outlined in \cite{susyds}.  First we make delta function localized operator valued measures, satisfying
$$\int\ \psi (\Omega , \Omega_0 ) q(\Omega ) = \phi q(\Omega_0 ) ,$$ where $q$ is an arbitrary measurable section of the chiral spinor bundle on the two sphere and $$[\phi , \phi^{\dagger} ]_+ = p > 0 .$$ Then we smear them with conformal Killing spinors (CKS),
$$Q_{\alpha} = \int\ \psi (\Omega , \Omega_0 )q_{\alpha} (\Omega ).$$  We find
$$[Q_{\alpha} , \bar{Q}_{\dot{\beta}} ]_+ = p (1, \Omega_0)_{\mu} \sigma^{\mu}_{\alpha\dot{\beta}} .$$  We require that the rest of the algebra be such that the SUSY representation content includes the gravity multiplet.  This is the list of kinematic compactifications of HST to minimal SUGRA in $4$ dimensions, somewhat analogous to the list of free string theories with the same SUSY algebra.  

We can now repeat this construction in block diagonal (meaning that the square matrices formed by spinor bilinears are block diagonal) matrices of size $N_i$, with all the $N_i \rightarrow\infty$ at fixed ratio.  We get a Fock space of massless supersymmetric particles in the limit.  Note that the energy is proportional to the sum of the block sizes.

Now suppose that $N$ is finite, and consider states satisfying the constraint
$$\psi_i^A | s \rangle = 0 $$ for a range of $0 < i < K \ll N $ and a range of $A$ of order $N$.   The states gotten by acting with generators $\psi_i^{A\dagger} $ , with $i,A$ in the range below $K$, on the state annihilated by all the $\psi$ operators, can be rearranged into various combinations of particles with total energy $K$, as above, as long as $K \gg 1$.  The entropy deficit of this set of states is of order $KN$.
Thus, if the Hamiltonian of the finite system is a small perturbation of the approximate SUSY Hamiltonian then eigenstates of this Hamiltonian with be distributed with Boltzmann probability with temperature proportional to $N^{-1}$, in the ensemble of all states of the system.

If we write 
$$H = P_0 + \frac{1}{N^2} {\rm Tr}\ F(\psi \psi^{\dagger}) , $$ where $F$ is a generic polynomial, then the eigenstates of $P_0$ will have lifetimes of order $1/N$, as a consequence of conventional large $N$ scaling.  For at least some choices of $F$, this Hamiltonian will be a fast scrambler and the dynamics of the system will be analogous to what we expect for a geodesic static observer in dS space.  The observer sees massless superparticles, for a time of order the dS radius, after which they get absorbed into a thermal system ``on the horizon".  For accelerated observers, at fixed static coordinate $r$, we instead write
$$H(r) = \sqrt{1 - \frac{r^2}{R^2}} P_0 + \frac{1}{N^2} {\rm Tr}\ F(\psi \psi^{\dagger}) . $$  This has the effect of blue-shifting the effective temperature felt by such an observer, in the way expected from semi-classical physics.

The reader who wants to learn more about the models sketched here should consult \cite{holounruh} and \cite{holoreview} when they appear.




\end{document}